\documentstyle[epsf]{cupconf}
\input {epsf.tex}
\begin{document}
\setcounter{part}{10}
\affiliation{$^1$ Inter-University Centre for Astronomy and
Astrophysics\\
Post Bag 4, Ganeshkhind, Pune 411 007, India\\
$^2$ Present Address: Department of Physics and Astronomy\\
UWCC, Post Box 913, Cardiff, CF2 3YB\\
U.K.
}
\title[Signal analysis]{Signal analysis of gravitational waves}
\author[B.S. Sathyaprakash]{B.S. Sathyaprakash$^{1,2}$}
\setcounter{page}{1}
\maketitle
\firstsection

\section{Introduction}
Signal analysis is an important component in the detection of gravitational
waves from astrophysical sources. The primary goal of signal analysis is
signal extraction or signal detection. There are several reasons why signal
extraction
will be a demanding exercise in searching for signatures of gravitational
radiation in the output of an interferometric detector. Firstly, an
interferometric detector is a wide band instrument, though in principle
it can be tuned to gain a larger sensitivity in a narrow band. Typically
the bandwidth is $\sim$ 1 kHz.
In addition to sampling useful data it is necessary to monitor
outputs from several channels continuously to assess the performance
of the detector. Consequently, the data output rate is expected to be
high $\sim$ 1 Mbyte s$^{-1}.$ Due to the high rate of data output it
is necessary to analyse the detector output online as otherwise data storage
requirements would be very expensive. Secondly,
gravitational wave antenna is more or less
an isotropic detector and does not point towards any particular direction.
This is an advantage in an important way since at once one is looking for
sources almost all over the sky. However, this calls for formidable amount
of data processing for sources that have to be integrated for a long duration,
such as a pulsar
emitting periodic gravitational waves, wherein it would be necessary to
correct for Doppler modulations of the signal wave form caused by
Earth's motion. The Doppler correction depends both on the location of the
source and frequency of the wave and current estimates show that
it would be computationally impossible to search for periodic sources all
over the sky and in a wide frequency band,
in a data train that lasts a couple of months (\cite{bfsles}).
Such Doppler corrections will be important even in the
case of quasi-periodic sources, such as a compact binary consisting of
two neutron stars or a neutron star and a white-dwarf, if the detector
is sensitive at very low ($<1$ Hz) frequencies. Thirdly, since there
are many candidate astrophysics sources---supernovae, pulsars, coalescing
binaries, to name a few---it is essential
to synchronously search for radiation from each of these sources.
Search algorithms for different sources are not all identical and hence
this causes extra burden on data analysis systems. Fourthly, looking for
stochastic background of gravitational radiation or for unknown sources,
requires cross correlation of two or more detector outputs. Since the light
travel time from one detector to another is significantly larger than the
time interval between consecutive data samples, correlations will have
to be carried out for several distinct overlaps of outputs from different
detectors.  Finally, even though, as in some cases, the signal wave form may be
known accurately,
the source parameters would be unknown and the data needs to be filtered
through several thousand templates. (For a review on data analysis problems
in gravitational wave detection see eg. Schutz (1989).)

When the detectors are built and start operating, it may very well
be that most of the analysis can be carried out just on a state-of-the-art
workstation of that time. However, certain searches---notably all-sky,
all-frequency search for pulsars which would be an important exercise towards
understanding the pulsar population in our Galaxy---appear to be
computationally
quite demanding even by the standard of computers expected to be available
towards the turn of the century (\cite{bfsles}). It is therefore essential
to work out
efficient data analysis algorithms in each case; efficient not only in
picking a weak signal from a noisy data but efficient in terms of computing
requirements.

The second aspect in signal analysis is the estimation of parameters
characterising a gravitational wave signal. After an initial detection
has been made it is often possible to deduce
the astrophysical quantities of the system emitting radiation, such
as the masses and spin angular momenta of stars in a binary system,
direction to a pulsar, etc.
Such an estimation can ultimately be used in testing theories of gravitation,
models of astrophysical systems and cosmology and so on.

In this lecture I will touch upon some aspects of signal analysis techniques
that are currently favourite. I will indicate a couple of alternatives in
the end but
the goal of this lecture is to give you an idea of what is involved in
signal processing. Signal analysis is expected to be a laborious exercise
only for pulsar searches and, to some extent, coalescing binaries and
stochastic background.
Aspects related to pulsar searches can be found in lectures by Schutz and
those related to stochastic background are dealt by Allen (1996);
in this
lecture I will only cover issues related to the detection and measurement
of coalescing binaries. Throughout this article we shall work with
geometrical units: $G=c=1.$

\section {Coalescing binary signal}

Highly evolved systems of compact binaries,
such as NS-NS, NS-BH~\footnote {NS=Neutron Star, BH=Black Hole.}, 
are amongst promising sources
of gravitational radiation for the planned laser interferometric
gravitational wave detectors.
As a binary system of  stars inspirals, due to radiation
reaction, the gravitational wave sweeps-up in amplitude and
frequency.
The resulting inspiral wave form has a characteristic power-law
spectrum $f^{-7/3}$ (\cite{kst87}) and is often called
the chirp or the coalescing binary wave form.
The wave form is worked out perturbatively
using post-Newtonian (henceforth PN) or post-Minkowskian expansion
and the lowest
order expression is called the Newtonian or the quadrupole wave form.
It turns out that for the purpose of detection of gravitational
waves from inspiralling binaries it is
sufficient to work with the so called {\it restricted} PN
wave form (\cite{3min}). In this approximation one incorporates
the PN corrections only to the phase of the
wave form working always with just the Newtonian amplitude.
Going beyond the restricted PN approximation is not
expected to change appreciably the magnitude of the statistical
errors in the parameter extraction, so the restricted PN
approximation can be used to estimate statistical errors as well. However,
in the post-detection analysis it is necessary to employ more
accurate templates since the use of just the restricted PN
wave form would give rise to some systematic errors. This is an
issue worth further exploration and we shall see one way of accomplishing
this. In this lecture, however, we shall always deal with the restricted
PN wave form.

As we shall see below all issues related to signal analysis can be
addressed in the Fourier domain.  In the stationary phase approximation
the Fourier transform of the restricted 2-PN \footnote {The notation
used is that $n$-PN corresponds to a term $v^{2n}$ beyond the quadrupole
term.}
chirp wave form (\cite{bdiww}) for positive frequencies is given by
\begin {equation}
\tilde h (f) = {\cal N} f^{-7/6} \exp \left [i\sum_{k=1}^6\psi_k(f)\lambda^k
- i {\pi \over 4} \right ].
\label {FT}
\end {equation}
For $f<0$ the Fourier transform is computed using the identity $\tilde
h(-f) = \tilde h^*(f)$ obeyed by real functions $h(t).$
In the above expression $\psi$'s are functions only of
$f$ and do not contain the parameters of the
wave form $\lambda_k,$
\begin {eqnarray}
\psi_1 & = & 2\pi f, \\
\psi_2 & = & -1, \\
\psi_3 & = & {6\pi f_a \over 5} \left ( {f\over f_a} \right )^{-5/3},\\
\psi_4 & = & 2\pi f_a \left ({f\over f_a}\right)^{-1},\\
\psi_5 & = & - 3\pi f_a \left ({f\over f_a}\right)^{-2/3},\\
\psi_6 & = & 6\pi f_a \left ({f\over f_a}\right)^{-1/3},
\label {FTphase}
\end {eqnarray}
$\cal N$ is a normalisation constant, $\lambda^k,$ $k=1,\ldots, 6,$ represent
the signal parameters
\begin {equation}
\lambda^k = \left \{t_C, \Phi_C, \tau_0, \tau_1, \tau_{1.5}, \tau_2 \right \},
\end {equation}
$t_C$ and $\Phi_C$ are the instant of coalescence of the binary and the phase
of the
wave form at that instant and
$\tau$'s are constants having dimensions of time and are called chirp times:
In terms of the total
mass $M$ of the two stars and the ratio of the total mass to reduced mass
$\eta=\mu/M$ the chirp times are
given by
\begin {equation}
\tau_0 = \frac {5} {256 \eta M^{5/3} (\pi f_a)^{8/3}},
\label {NCT}
\end {equation}
\begin {equation}
\tau_1 = \frac {5} {192(\pi f_a)^2 \eta M}
\left ({743\over 336} + {11\over 4} \eta \right ),
\end {equation}
\begin {equation}
\tau_{1.5} = \frac {\pi}  {8 (\pi f_a)^{5/3} \eta M^{2/3}} 
\end {equation}
and
\begin {equation}
\tau_2 = \frac {5} {128 \left (\pi f_a \right )^{4/3}\eta M^{1/3}} 
\left [{3058673 \over 1016064} + {5429\over 1008} \eta + 
{617\over 144}\eta^2 \right].
\end {equation}
The chirp times depend on the two masses of the stars and the lower frequency
cutoff $f_a$ of the detector and hence all are not independent.
The physical significance of $\tau$'s is that they contribute to the total
inspiral time of the binary starting at a time when the gravitational wave
frequency is $f_a$; $\tau_0$ is the
Newtonian contribution and others are the various PN
contributions, the total chirp time being $\tau_0+\tau_1-\tau_{1.5}+\tau_2.$

\setcounter{equation}0
\section{Detection}

In this Section we cover aspects related to the detection of
gravitational waves with the aid of matched filtering.
First, we shall focus attention on the
{\it detection statistic} and define the {\it scalar product}
of two signal wave forms. The latter facilitates a quick derivation
of an expression for the {\it optimal Weiner filter} and
the notion of the {\it ambiguity function}
which is a very powerful tool to deal with aspects of detection
and estimation.

\subsection {Detection statistic}

Weiner filtering or matched filtering is a data analysis
technique that efficiently searches for a signal of known shape 
(\cite{cwh68}).
The method consists in correlating the raw
output of a detector with a wave form, variously known as a
template or a filter. Given a signal $h(t;\vec\lambda),$
where $\vec \lambda=\{\lambda_k | \ k=1,\ldots,N_s\}$ denotes
the $N_s$ signal parameters, buried in noisy
data $n(t),$ the task is to find an `optimal' filter $q(t;\vec\mu)$ that would
produce, on the average, the best possible
signal-to-noise ratio (SNR) to be defined below.
Here $\vec\mu =\{\mu_k | \ k=1,\ldots,N_f\}$ denotes the $N_f$
filter parameters.
Let us denote by $o(t)$ the output of the detector:
\begin {equation}
o(t) = h(t; \vec\lambda) + n(t).
\end {equation}
The correlation $c$ of the filter with
the detector output is defined as
\begin {equation}
c(\tau;\vec\mu) \equiv \int_{-\infty}^{\infty} o(t) q(t+\tau; \vec\mu) dt
= \int_{-\infty}^{\infty} \tilde o(f) \tilde q^*(f; \vec\mu)
\exp\left (2\pi i f \tau\right ) df
\end {equation}
where $\tau$ is the lag parameter,
$^*$ denotes complex conjugation,
$\tilde{}$ denotes the Fourier transform of the quantity underneath
($\tilde a(f) \equiv \int_{-\infty}^\infty a(t) \exp \left ( -2 \pi i f t
\right ) dt$)
and the second equality follows from Parcevel's theorem.
Now, $n(t)$ being a random variable, so is $c(\tau; \vec \mu)$ and a
decision about the presence or absence of a signal is
made by setting a threshold on the detection statistic of matched
filtering which is the SNR $\rho$ defined as
\begin {equation}
\rho \equiv \max_{\tau,\vec\mu}
{\overline {c(\tau;\vec\mu)} \over
\sqrt {\overline {\left [c(\tau;\vec\mu) - \overline {c(\tau;\vec\mu)}\right
]^2}}},
\end {equation}
where an overbar denotes ensemble average of the quantity underneath it over
different realisations of the detector noise. The noise is assumed to be
a Gaussian random process with zero mean and real, symmetric,
two-sided power spectral density
$S_h(f)$ defined by
\begin {equation}
\overline {\tilde n(f) \tilde n^*(f')} = S_h(f) \delta(f-f').
\end {equation}
Before proceeding further let us define the scalar product of
functions which plays a crucial role in signal analysis. Given
two functions $a(t)$ and $b(t)$ their scalar product is defined
as
\begin {equation}
\left < a| b \right > \equiv \int_{-\infty}^\infty
{df \over S_h(f)} \tilde a(f) \tilde b^*(f).
\end {equation}
The scalar product is real and positive definite owing to the
properties of the noise power spectral density and Fourier
transform of real functions.
A wave form is said to be {\it normalised} if its norm, computed using
the above definition of the scalar product, is equal to unity:
$\left < h, h \right >^{1/2}  = 1.$
With the aid of this notation the statistic $\rho$ takes the form
\begin {equation}
\rho = {\left <h, S_h q\right > \over\left <S_h q, S_h q\right > }.
\end {equation}
>From this it is clear that the template $q$ that obtains the
maximum value of $\rho$ is simply
$\tilde q(f)= \gamma \tilde h(f) / S_h(f)$ where $\gamma$ is
an arbitrary constant.
Thus, in the Fourier domain an optimal filter
is nothing but the signal weighted down by the noise power
spectral density.  In order to decide whether or not a signal
is present the detector output
is filtered through a bank of templates which are chosen
over the entire parameter space of the template and the optimal SNR is
$\rho=\left <h|h\right>^{1/2}.$
A template whose parameters are exactly matched
with those of a signal enhances the SNR
in proportion to the square-root of the number of
cycles that the signal spends in the detector output,
as opposed to the case when the shape of the wave form is
not known a priori and all
that can be done is to pre-bandpass filter the detector output
to the frequency band where the
signal is assumed to lie, and to then look at the SNR for each
data point in the time domain individually (\cite{bfs89}).
For an interferometric detector, such as the LIGO (\cite{LIGO})
or the VIRGO (\cite{VIRGO}) ,
operating with a lower frequency cutoff $\sim$~40~Hz and
an upper cutoff $\sim$~1~kHz,
this means an amplification in the SNR~$\sim$~30-40
for a typical binary. This
enhancement in the SNR not only increases the number of
detectable events but, more importantly, it also allows a more accurate
determination of signal parameters---the error in the
estimation of a parameter being inversely proportional
to the SNR.

\subsection {Ambiguity function}

In order to take full advantage of matched filtering
it is essential that the inspiral wave form,
and in particular the evolution of its phase, be
known to a very high degree of accuracy (\cite{3min}). A mismatch in the
phases of the template and the signal can severely reduce the SNR;
even when the template and the signal go out of phase
by one cycle in $10^4$ the SNR could reduce by as much as 10\%.
This is a positive aspect of Weiner filtering
since the statistic will not pick out any spurious
chirp-like signals present in the detector output.
However, the problem is that we will not know a priori what the
signal parameters are and consequently, the detector output
needs to be filtered through a large number of templates each
corresponding to a particular set of ``test'' parameters.
Thus, we have to finely sample the parameter space in searching for
a signal.  The number of search templates needed to cover the
astrophysically relevant range of the parameter space
depends primarily on the effective dimensionality of the parameter
space. Ambiguity function, well known in statistical theory of
signal detection (\cite{cwh68}), is a very powerful tool in signal analysis:
It helps to assess the number of
templates required to span the parameter space of the
signal, to make estimates of variances and covariances involved
in the measurement of various parameters, to compute biases introduced in
using a wrong family of templates, etc. Later in this Section we will
see how the ambiguity function can be used to compute the number
of templates.

The ambiguity function is defined as the scalar product of
two normalised wave forms
$h(t; \vec \lambda)$ and $g (t; \vec \mu)$ which differ in all their
parameter values, i.e., $\vec \lambda$ being in general different from $\vec
\mu:$
\begin {equation}
{\cal A} (\vec \lambda, \vec \mu) \equiv \left <h (\vec \lambda), g (\vec \mu)
\right >.
\label {defcorr}
\end {equation}
Since the wave forms are of unit norm
${\cal A} (\vec\lambda, \vec\mu) = 1,$ if $\vec\lambda = \vec\mu$ and
${\cal A} (\vec\lambda, \vec\mu) < 1,$ if $\vec\lambda \ne \vec\mu.$
Here $\vec\lambda$ can be thought of
as the parameters of a signal while $\vec\mu$ those of
a template. With this interpretation ${\cal A} (\vec\lambda, \vec\mu)$ is
the SNR obtained using
a template that is not necessarily matched on to the signal.
Keeping the filter parameters $\vec\mu$ fixed if we vary the signal
parameters $\vec\lambda$ the ambiguity function is a function of
$N_s$ signal parameters
giving the SNR obtained by the template for signals of different
values of their parameters.  The region in the signal parameter space
for which a given
template obtains SNRs larger than a certain value (sometimes
called the {\it minimal match} (\cite{bjo96})) is the {\it span}
of that template and the templates should be so chosen that together
they span the entire signal parameter space of interest with
the least overlap of
one other's spans. One can equally well interpret the ambiguity function as
the SNR obtained for a given signal by filters of different parameter values
and in this case the ambiguity function is a function of
$N_f,$ rather than $N_s,$ variables (see below the Section on biases
for an application of this interpretation). Of course, in its entirety
the ambiguity function is really a function of $N_s + N_f$ variables.

It is important to
note that in the definition of the ambiguity function there is
no need that the functional form of the template be that of a
signal; the definition holds good for any signal-template
pair of wave forms. Moreover,
the set of $N_f$ template parameters $\vec \mu$ need not be identical
(and usually aren't) to the set of $N_s$ parameters $\vec \lambda$
characterising the signal.
For instance, a binary can be characterised
by a large number of parameters, such as the masses, spins, eccentricity of the
orbit, etc., while we may take as a model wave form for the purpose
of filtering the one only involving the masses. This is indeed an
issue where substantial work is called for: What are all the
physical effects to be considered
so as not to miss out a binary wave form from our search?

It is thus possible to render a wider
interpretation to the ambiguity function, the one that is conceptually
more appropriate and useful in the context of gravitational waves.
Such an interpretation is based on the geometrical view point of signal
analysis which is briefly the following:  A wave form $h(t;\vec\lambda)$
with a specific
set of values for its parameters can be thought of as a {\it signal vector.}
As its $N_s$ parameters are varied the  signal vector spans a $N_s$-dimensional
space in the underlying infinite-dimensional function vector space.
The former has a manifold structure, the parameters of the signal constituting
a coordinate system and the metric defined using the
scalar product introduced earlier.  A given template $g(t;\vec\mu)$ is itself
a vector but may or may not be a member of the signal space. The set of
all template wave forms obtained by varying the $N_f$ filter parameters
span an $N_f$-dimensional space which again has manifold structure.
By choice the template manifold is of a dimension lower than the signal
manifold. Now, the ambiguity function, viewed as a function of its arguments
($\vec \lambda$ and $\vec \mu$),
gives the nearest distance between different points
on the filter manifold from points
on the signal manifold, distance being measured using the scalar product.
In the context of gravitational waves $h(t;\vec\lambda)$ is the exact
general relativistic wave form emitted by a binary while
the template family, as of now, is the 2-PN
corrected chirp. Of course, in this case we cannot compute the ambiguity
function
since the exact wave form is not known (if it had been known, we would use that
in our family of templates instead of the 2-PN wave form).
However, it is possible to obtain ambiguity function similar to this but
in the case of
a test particle orbiting a Schwarzschild black hole. In the latter case
the exact phase
evolution of the wave form is known numerically and
approximations to the phase evolution are known analytically up to
4-PN order. Such a study would then
help us to make intelligent guesses about what will happen when both
stars have comparable masses which in turn can be used to assess the
effectiveness of a PN filter.
The realisation that filters need not have shapes identical to
a signal has led to the proposition
that it might be worthwhile to consider templates that do not belong
to any of the PN manifolds but lie outside of them (\cite{bsd96}).

To gauge the usefulness of various PN
template families one can compute the ambiguity function by taking
$h(t;\vec\lambda)$ to be the 2-PN corrected signal
and consider template families of different PN orders
starting with the quadrupole wave form. Though this is not as good a
method as suggested in the preceding paragraph it does
address some of the issues.
Such an attempt was first made by Balasubramanian  \&  Dhurandhar 
(1994)
and Kokkotas, Krolak \&  Tsegas (1994)
and was investigated more formally and exhaustively by
Apostolatos (1995).
The result of all this analysis is that the 1-PN
template families are inadequate in the detection of gravitational waves
and that one has to work with at least 1.5-PN wave forms (\cite{taa95}).
A more reliable understanding is likely to come
by studying the test particle case mentioned above.

Let us employ the ambiguity function to find how many templates are
needed to span the range of astrophysical parameters using different
families of templates. For brevity we shall only deal with the Newtonian
and the 1-PN template families and simply quote results in
other cases.
In general, as indicated by its arguments ${\cal A} (\vec\lambda, \vec\mu)$
depends on the individual values of the parameters
both of the signal and the template. In what follows we will first see
that in the case of restricted PN chirps to 1-PN order
the ambiguity function only depends on the absolute difference
in the parameter values $|\lambda_k-\mu_k|$ provided the template
and the filter are both chirps to the same PN order. Even at higher PN
orders there is only a weak dependence on the absolute values of 
the parameters. Geometrically, this means that the signal manifold at
restricted 1-PN order is flat and that higher PN order signal 
manifolds are only slightly curved.
Secondly, we will see that a template of a given total chirp time
obtains roughly the same SNR for signals of the same total chirp time
though their Newtonian and PN chirp times may be
different from that of the template.
The former of these two results implies
uniformity in the spacing of filters (\cite{bssd91}; \cite{sdbs94}) while the latter
result facilitates a massive reduction in the number of
templates required in spanning the parameter space since instead
of constructing filters separately for each of the
Newtonian and PN chirp times we can construct filters
simply for the total chirp time.

Making use of the expression for the Fourier transform given in eq. (\ref {FT})
when the signal and templates belong to the same PN order
we have
\begin {equation}
{\cal A} (\lambda_k,\mu_k) = 2 {\cal N}^2\int_0^\infty
{df\ f^{-7/3}\over S_n(f)} \cos
\left [ \sum_{k=1}^n\psi_k (f) \delta_k \right ]
= {\cal A} (\delta_k)
\label {corr1}
\end {equation}
where $\delta_k = |\lambda_k - \mu_k|$ and $n-3$ ($n\ge 3$)is the 
PN order of the wave forms. We see that the
ambiguity function is independent of the individual chirp times
of the signal and the template: {\it For all signal-template
pairs that have the same differences in times of arrival, phases,
and chirp times one obtains the same value for the ambiguity function}
(\cite{bss94}; \cite{bsd96})
\footnote {This is only an approximate property of the ambiguity function
since the result relies on the accuracy of the stationary phase method
employed in computing the Fourier transform of the chirp. Moreover,
real wave forms are shutoff after the signal has reached a frequency
corresponding to the last stable circular orbit of the
binary which occurs when the distance between the two stars is $6M.$
However, results concerning the ambiguity function hold good only when
all wave forms have the same upper frequency cutoff.}.
Consequently, constancy of the distance,
measured using the scalar product,
between two nearest neighbour filters translates into the
constancy of the {\it distance}, measured using the difference
in their parameter values. As a concrete example let
us take both the signal and filters to be the Newtonian wave form.
In this case there is only one parameter for which filters need to
be explicitly constructed, namely the Newtonian chirp time.
The ambiguity function obtained by using the noise power spectral
density expected in the initial LIGO interferometer (\cite{sfdc93}; \cite{ccef94}) and
keeping the chirp time of the template constant at 4 s while varying
the signal parameters in the range of 3.8 to 4.2 s, \footnote {
Newtonian chirp time 4 s corresponds to an equal mass binary
of total mass $8.5$ $M_\odot$ assuming a lower
frequency cutoff of 40 Hz as in the case of initial LIGO.}
is plotted in Fig.~1 as
solid line. Identical curves are obtained irrespective what value
we choose for the chirp time of the template confirming the claim
made above.

\begin{figure}
\centerline{
\epsfxsize=4. in \epsfbox{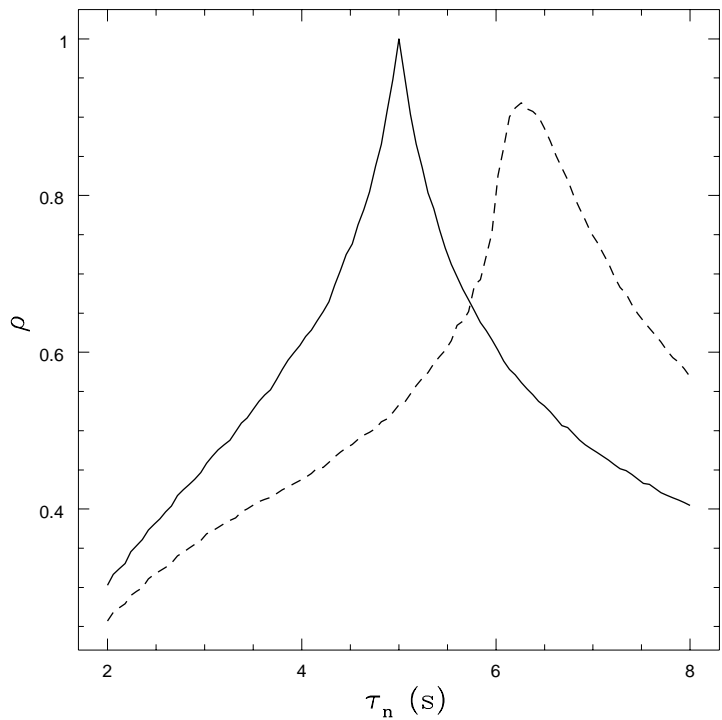} 
}
\caption{Plot showing the behaviour of the ambiguity function
when both the signal and filter are quadrupole wave forms (solid line)
and when the signal is a 1-PN corrected wave form and
the templates are Newtonian (dashed line).}
\label{Fig1}
\end{figure}

Now, if we want to search for all binaries with the mass
of each star in the range $[0.2, 30] M_\odot$ then we need to span the range
$[\sim 0, 660]$~s of the Newtonian chirp time. By drawing a horizontal
line at, say ${\cal A} = 0.97,$ we get the span of a template to be
$0.030$~s which gives the total number of filters to be about 
$2.2 \times 10^4.$
Estimates of the number of filters required to span the parameter
space of PN signals (for which the dimensionality of the
parameter space is two---masses of the two stars, or any two of the
chirp times introduced earlier, or the total and the reduced mass of the
system, any of these pairs of parameters serve to characterise the signal)
are only an order of magnitude larger ($\sim 2.4 \times 10^5$) (\cite{bjo96})
and not $10^8$ as one would have expected from the dimensionality of the
parameter space. Even at 2-PN order the number of templates is expected
to remain the same, since there are no new parameters,
except that the spacing between templates will no more be uniform.

It is worth pointing out the relation between the number of filters
and the bandwidth of a detector.
Keeping the sensitivity constant if we increase the bandwidth then
we have a wider frequency range to compare two wave forms and hence
the correlation between two signals of differing parameters would grow
smaller. Decreasing the bandwidth has just the opposite effect.
Concerning parameter estimation, at any given SNR,
greater the bandwidth larger will be the accuracy with which parameters
can be measured. It seems from this that if we want a better parameter
estimation then we have to cope with a larger number of templates in
our search (see Balasubramanian {\sl et al.} (1996) 
for more details on this and related
issues).

\section {Estimation}

After a detection is made the next step in data analysis is estimation
of parameters characterising an event and provide possible error bounds
on the measured values. We will discuss aspects related to estimation in this
section.

\subsection {Covariance matrix and error bounds}

In our search for a chirp signal in the output of an interferometric
detector we use a discrete, rather than a continuous, family of
templates. The spacing between templates could be quite large, as in the
case of a hierarchical search, or small, as in the case of post-detection
analysis. The parameters of the template that obtains
the maximum SNR gives us a maximum likelihood estimate as such a template
would also maximise what is called the maximum likelihood ratio. These
estimates are most unlikely the actual parameters of the signal; the true
parameters are expected to lie within an ellipsoid of $N_s$ dimensions
at a certain level of confidence---the volume of the ellipsoid increasing
with the level of confidence. The axes of the ellipsoid are the one
sigma uncertainties in the estimation of parameters and the confidence
level corresponding to a one sigma uncertainty is 67\%. The errors
(i.e. one sigma uncertainties) in the various parameters are given by
the square root of the diagonal elements of the covariance matrix
$C_{ij}.$ The latter is the inverse of the Fisher information
matrix $\Gamma_{ij}$  well known in statistical theory of signal detection
(\cite{cwh68}) given by
\begin {equation}
\Gamma_{ij} = \left < {\partial h(\vec \lambda) \over \partial \lambda_i},
{\partial h(\vec \lambda) \over \partial \lambda_j} \right >,
\ \ \ C_{ij} = \left (\Gamma \right )^{-1}_{ij}.
\end {equation}
Bounds on the estimation computed using the covariance matrix are called
Cramer-Rao bounds. These are not very tight bounds on estimation; they
are the minimum uncertainty one should expect for the various parameters.
They are based on local analysis and do not take into consideration
the effect of distant points in the parameter space on the error computed
at a given point.
The errors are typically much larger than that predicted by the
covariance matrix and have to be computed by other,
more involved, methods. We shall see one such numerical method in the
next Section.  Cramer-Rao bounds fall off as inverse of the SNR while
actual errors do not necessarily follow this behaviour. One usefulness
of the Cramer-Rao bounds is that they are asymptotically valid in the limit
of high SNR and hence serve as a basis to test all other estimates of
errors. Covariance matrix based errors in the estimation of the Newtonian
and PN chirp times ($\tau_0$ and $\tau_1$) and the instant
of coalescence $t_C$ are listed in Table I for different values of the
SNR\footnote {For brevity we have not included the phase parameter
$\Phi_C$.}. Due to the
fact that the scalar product of two chirps of the same PN family
is independent of the absolute values of the parameters, it turns out
that the errors in chirp times and the instant of coalescence are independent
of which binary we are looking at. This is not so if we were to use as
our parameters the masses of the stars or the reduced and total mass.
Of course, the relative error in chirp times will be larger, for a given
SNR, in the case of higher mass binaries since the latter last for a smaller
duration. From Table I we see that a chirp detected at an SNR of 10
is expected to lie, according to covariance matrix estimates,
within an ellipsoid of dimensions
$45 \times 26 \times 0.26 $~ms$^3$ at a confidence level of 67\% and within
an ellipsoid of 8 times larger volume at a confidence level of
95 \%, and so on. (Estimates of errors for various PN signals can be
found in (\cite{sfdc93}; \cite{ccef94}; \cite{lbbs94}; \cite{epcw95}; \cite{bsd96}) and a quick theory
of estimation is given in (\cite{lsf92}; \cite{ccef94}; \cite{pjak95}).)

\begin{table}
\begin {center}
\label{pntab}
\caption{Errors in parameter estimation computed using the covariance matrix
$(\sigma_{\tau_0},$ $\sigma_{\tau_1},$ $\sigma_{t_C})$ are listed at various
SNRs. Corresponding errors found with the aid of Monte Carlo simulations
$(\Sigma_{\tau_0},$ $\Sigma_{\tau_1},$ $\Sigma_{t_C})$
are quoted in brackets.  All values are quoted in ms.}
\begin{tabular}{ccccc}
\\
\hline
\hline
&$\rho=10.0$  &  $\rho=15.0$  &  $\rho=20.0$ & $\rho=25.0$ \\
\hline
$\sigma_{\tau_0}$ $(\Sigma_{\tau_0})$
& 45 (135) & 30 (65) & 23 (30) & 18 (19) \\
$\sigma_{\tau_1}$ $(\Sigma_{\tau_1})$
& 26  (65)  & 17 (33) & 13 (16) & 10  (11)  \\
$\sigma_{t_C}$  $(\Sigma_{t_C})$
& 0.26  (0.54) & 0.17 (0.30) & 0.13 (0.17) & 0.10  (0.10) \\
\hline
\end{tabular}
\end {center}
\end{table}

\subsection {Biases in estimation}

Ambiguity function can also be used to estimate biases.
To do this we need the alternate interpretation of the ambiguity function
namely, that it gives the SNR obtained for a given signal by different filters.
Thus, we keep the parameters of the signal to be constant and find which
amongst all filters obtains the maximum SNR.
Let me illustrate this with a simple example.
Suppose the filters are Newtonian wave forms and the signal is the
2-PN wave form. The ambiguity function obtained in this
case, plotted as a dashed curve in Fig.~1, is not a symmetric curve and it does
not any more possess the nice properties of being independent of the
absolute values of the parameters.
Moreover, the maximum has shifted: The maximum SNR is
obtained not by a filter that
matches the Newtonian chirp time of the signal but by some other filter
whose chirp time is different from that of the signal. If the real world
had consisted of 2-PN wave forms and we had used Newtonian
wave forms we would not only miss a large number of events
but our parameter estimation would also come out wrong.
In the present case there is 15\% loss in the SNR and 
$-40$~ms  bias in the estimation of Newtonian chirp time.
In reality the wave form present in the detector is the fully general
relativistic wave form with all the non-linearities but the template
that we hope to use will be some high order PN approximation
to it and consequently we are bound to make systematic errors in estimation.
In the case when a lower order PN template family is
used in detecting a higher PN signal,
filters seem to match the instant of coalescence $t_C$ to
an accuracy better than any other parameter (\cite{bsd95}).
This is likely to be true even when PN templates are used
to pick out the actual signal buried in our detector outputs.
Based on this result we expect
that $t_C$  will be measured to a greater accuracy than any other parameter
and it would be sensible to use this, or any other parameter that is
determined best, in our tests of theories and models.

\subsection {Monte Carlo estimation of parameters}

As mentioned in the previous Section Cramer-Rao bounds are only a
lower limit on the errors expected and realistic errors are much
larger than this. One brute force way of estimating the errors is
to carry out numerical simulations mimicking the actual detection
process. An advantage of numerical simulations is that unlike in the
derivation of the covariance matrix they make no assumption about
the behaviour of the filtered noise distribution function. Ensemble
averages that are required to make estimates are achieved by performing
a large number of simulations corresponding to different
realisations of the expected detector noise in each numerical experiment.
This is equivalent to having a large number of detectors and upon filtering
and maximising each detector output over the space of template wave forms
gives a certain measured set of values
of the parameters. The average of the measured values provides
estimates of the parameters and variances in the measured values give
typical errors involved in any one measurement. A simulation such as this
was carried out to assess the degree of accuracy of the covariance matrix
(\cite{bsd95}; \cite{bsd96}).
Results from this study show that the covariance matrix underestimates
the errors of various parameters by factors of 2--3 (cf. Table I) 
at an SNR of 10. However,
at SNRs larger than $\sim 25,$ errors obtained by Monte Carlo simulations
agree with those predicted by covariance matrix calculations.
There has been an effort to understand the discrepancy between the two
results (\cite{vd96}) which clearly indicates that the covariance matrix
is a poor descriptor of errors and better greatest lower bounds on errors
are in order.

Even though realistic errors are a factor of 2--3 larger
than those predicted by the covariance matrix in absolute terms
they are still quite small so that it would be possible to put theories of
gravitation and models
of cosmology to test and constrain them at high confidence levels.
For instance, it has been shown that presence of non-linear tails of
gravitational waves, occurring due to the scattering and subsequent
re-emission of radiation off the curved Schwarzschild spacetime of the
binary, can be detected in the case of BH-BH coalescences that have
an SNR $\sim$ 50 or larger (\cite{lbbs95}). It has also been suggested that
statistics based on catalouges of coalescing binary events are potential tools
to measure cosmological parameters and test cosmological models (\cite{lsf96}).

\section {Summary and future}

In this article we have discussed issues concerning the detection and
measurement of gravitational radiation from coalescing compact binary
systems. The fully general relativistic wave form that will be
buried in the data stream is not known to us and
we can only hope to use an approximate wave form that is presently
known to an accuracy $(v/c)^{4}$ beyond the Newtonian order.
Efforts are imminent to gauge the PN
level up to which it is essential to know the wave form in the
finite reduced mass case by studying the test particle case where
both the exact and the approximate solutions are known (\cite{dis96}). 
There are other effects such as the eccentricity-induced or
spin-induced modulations which may
affect the wave form so much as to make it inefficient in picking
out a true coalescing binary signal. It is important to gauge these
effects.

A considerable amount of work has been done in determining the
number of templates required to span the signal parameter space
and present compute resources seem
adequate to filter chirp signals online. However, not all ideas
concerning the choice of filters are explored yet (\cite{bsd96}).
Moreover, one has to still take into account physical effects such
as spin-orbit and spin-spin coupling, last stable orbit etc., in computing 
the ambiguity function.

We have seen how biases are introduced in the measurement of parameters
and how the ambiguity function may be employed to compute these biases but
much needs to be done to make concrete estimates of the biases.

All efforts so far have concentrated on the study of matched filtering
as a tool to dig out chirps out of noise. However,
there are a number of other detection algorithms---chirplets,
adaptive filters, periodograms, Smith's algorithm, etc.---some of
which may be more ideally suited to the detection of chirps
than Winer filtering is. Further study in this direction is needed
to make our data analysis systems more efficient.

\begin{acknowledgments}
It is a pleasure to thank R. Balasubramanian, L. Blanchet,
T. Damour, S.V. Dhurandhar, L.S. Finn, A. Gopakumar, B.R. Iyer,
D. Nicholson and B.F. Schutz for many useful discussions and
Jean-Alain Marck and Jean-Pierre Lasota
for organising a most stimulating School and for hospitality.
I am indebted to Balasubramanian for comments on an earlier version
of this article.
\end{acknowledgments}

\end{document}